\documentclass[a4paper,11pt]{article}
\pdfoutput=1 % if your are submitting a pdflatex (i.e. if you have
\usepackage{jinstpub} % for details on the use of the package, please
\title{A LN\textsubscript{2} Based Cooling System for a Next Generation Liquid Xenon Dark Matter Detector}

\author[a,b,1]{K.L. Giboni,\note{Corresponding author.}}
\author[a,b]{P. Juyal,}
\author[c]{E. Aprile,}
\author[c]{Y. Zhang,}
\author[d]{and J. Naganoma}

\affiliation[a]{INPAC and School of Physics and Astronomy, Shanghai Jiao Tong University,\\Shanghai 200240, China}
\affiliation[b]{Shanghai Laboratory for Particle and Cosmology,\\Shanghai 200240, China}
\affiliation[c]{Columbia Astrophysics Lab and Physics Department, Columbia University,\\New York, NY 10027, USA}
\affiliation[d]{Department of Physics and Astronomy, Rice University,\\Houston, TX 77005, USA}

\emailAdd{kgiboni@sjtu.edu.cn}

\abstract{In recent years cooling technology for Liquid Xenon (LXe) detectors has advanced driven by the development of Dark Matter (DM) detectors with target mass in the 100 -- 1,000 kg range. The next generation of DM detectors based on LXe will be in the 50,000 kg (50 t) range requiring more than 1 kW of cooling power. Most of the prior cooling methods become impractical at this level. For cooling a 50 t scale LXe detector, a method is proposed in which Liquid Nitrogen (LN\textsubscript{2}) in a small local reservoir cools the xenon gas via a cold finger. The cold finger incorporates a heating unit to provide temperature regulation.

The proposed cooling method is simple, reliable, and suitable for the required long-term operation for a rare event search. The device can be easily integrated into present cooling systems, e.g. the 'Cooling Bus' employed for the PandaX I and II experiments. It is still possible to cool indirectly with no part of the cooling or temperature control system getting in direct contact with the clean xenon in the detector. Also the cooling device can be mounted at a large distance, i.e. the detector is cooled remotely from a distance of 5 -- 10 m. The method was tested in a laboratory setup at Columbia University to carry out different measurements with a small LXe detector and behaved exactly as predicted.}

\keywords{Noble liquid detectors (scintillation, ionization, double-phase); Dark Matter detectors (WIMPs, axions, etc.); Large detector systems for particle and astroparticle physics; Very low-energy charged particle detectors; Time projection chambers; Cryogenics; Detector cooling and thermo-stabilization}

\begin{document}
\maketitle
\flushbottom

\section{Introduction}
\label{sec: Introduction}

In recent years progress in Liquid Xenon (LXe) detector technology has been driven by the search for Dark Matter (DM) in the form of Weakly Interacting Massive Particles (WIMPs). Despite the increase in target mass, from a few kilograms to several tons~\cite{1,2}, and the superior sensitivity reached by LXe based searches, WIMPs remain undetected. We are now entering the era of 5 -- 10 t detectors, with XENONnT~\cite{3}, LZ~\cite{4}, and PandaX IV~\cite{5}. To either confirm and increase the statistical significance of a detection, or to continue to explore the interaction cross section down to the level where neutrinos become an irreducible background, a LXe experiment at the 50 t scale is under study~\cite{6}. For a good overview of LXe detectors for dark matter search and other applications see Ref.~\cite{7}.

The cross sections of the rare interactions to be observed by such ultimate LXe detector are so small that even with a massive target one must take data continuously over periods of multiple years. The operating conditions must be kept constant for such long periods since any change requires a new sequence of calibration runs and possibly an additional analysis effort. This demands a cooling system which requires as few as possible interventions during such long run times. The cooling must be ultimately reliable, easy to operate with good stability on both short and long timescales. Furthermore not only the detector but also the cooling system must not compromise the low background requirement essential for rare event searches.

With the exception of the LUX detector, all DM detectors operated by the XMASS~\cite{8}, XENON~\cite{1,9,10} and PandaX~\cite{11} collaborations were cooled by Pulse Tube Refrigerators (PTRs)~\cite{12}. This kind of refrigerator has proven to be ideal for these low-background rare event searches --- easy to operate and control, and highly reliable. The use of a PTR also allowed remote cooling of detectors mounted in a water shield, at a distance exceeding 5 m. However, the cooling power of commercially available PTRs is limited to about 200 W. The power can be augmented by using several PTRs in parallel, but using more than two PTRs might be impractical and unreasonable because of the high costs of the units and also because of the high power consumption of the helium compressors.

There are not many alternative cooling methods for LXe. The solution we propose here seems to be the most practical and cost-effective, while fulfilling all requirements. The solution resembles a PTR cooling unit, but the PTR is replaced by a Liquid Nitrogen (LN\textsubscript{2}) reservoir. We combine the ease of operation and the convenience of PTR cooling with the elevated power of a LN\textsubscript{2} system. This method was tested at Columbia University in a small setup within the framework of the XENON detector development program several years ago. These tests were successful, but the cooling system was never described in a publication.

\section{Overview of Cooling Methods for LXe Detectors}
\label{sec: Overview of Cooling Methods}

Originally the necessity to cool xenon gas to realize a LXe radiation detector was considered a challenge. The temperature range from freezing to boiling point of xenon is very small, $-$112$^{\circ}$C to $-$108$^{\circ}$C at ambient pressure. One can moderately alleviate the problem by moving away from the Triple Point, i.e. choosing a higher operating pressure. The boiling temperature increases much faster then the freezing one. Typical operating temperatures are therefore around $-$95$^{\circ}$C, and the pressure is around 1.5 barG. Still one needs a tight regulation of the cooling power especially for dual phase detectors whose proportional amplification gain is pressure sensitive and varies substantially with changes in liquid level. The usual cooling media in a physics lab are not well suited for this temperature range. Dry-Ice is not cold enough, and LN\textsubscript{2} is far too cold. The system will be far from a thermal equilibrium, and some portions of the xenon might freeze. If there are no counter measures the active volume of the detector could convert into a solid ice block, most likely destroying the delicate electrode structure. The phenomenon of freezing can be easily observed by monitoring the pressure. When freezing starts the pressure will rapidly approach the low vapor pressure of xenon ice. In the sections below we briefly review the different cooling methods used for LXe detectors, to the best of our knowledge.

\subsection{Thermosyphon}

A thermosyphon is a closed system with three parts, the evaporator, the connecting pipe, and the condenser. LN\textsubscript{2} from a reservoir cools the condenser on one end of the pipe. Here gaseous nitrogen (GN\textsubscript{2}) under a controlled pressure is liquefied and runs to the other end propelled by gravity. At this end, the LN\textsubscript{2} boils off and cools the so-called evaporator, by the latent heat of the phase transition. The nitrogen gas from the evaporator finally is guided back towards the condenser through the same pipe. The system is a closed loop with heat being extracted from the evaporator and deposited in the condenser. The condenser has to be mounted on top of the detector. Of course, the nitrogen loop has to be hermetically sealed, and it has to be well insulated. The LUX experiment~\cite{13} used an arrangement of three thermosyphons, one on the bottom of the LXe detector and two on top, connected to a thermal shield. The thermal flow of the loop, and thus the cooling power, could be regulated by the filling pressure of the gaseous nitrogen in the connecting pipe.

Thermosyphons~\cite{14} up to 1 kW were tested, but probably this is not a limit. The units were engineered for highest efficiency with heat losses reduced to a minimum. The operating problems with some of the thermosyphons mentioned in the original publication are most likely solved in the meantime. Thermosyphons are an option for very large LXe detectors, but their design is more complicated than the solution described in Sec.~\ref{sec: LN2 with Cold Finger}.

\subsection{Dry-Ice - Freon}
In early R\&D with small LXe ionization chambers~\cite{15,16} an open bath of Freon at Dry-Ice temperature was used for cooling. At $-$78$^{\circ}$C the xenon liquefies at a pressure of about 5 bar. The cooling is of course very stable as long as Dry-Ice blocks remain in the bath. The high pressure is at the limit of the specifications for all the photomultiplier tubes (PMTs) which are used for DM searches. Only recently a 'high pressure' version of the Hamamatsu metal-channel PMT R8520~\cite{17} with 10 bar became available. Still the high pressure is a challenge. Naturally, a pressure vessel at 5 bar requires much thicker walls than at 1 bar. The additional material of the walls adds to the radioactive background of the detector.

Finally, environmental considerations and the increased costs of Freon strongly disfavored this method which was then replaced by the one described below.

\subsection{Alcohol - LN\textsubscript{2} Mixture}

To cool small size LXe detectors~\cite{7} the Dry-Ice - Freon bath was replaced by an open bath with an alcohol - LN\textsubscript{2} mixture. The detector vessel is immersed into an open-mouth dewar filled with ethyl alcohol. LN\textsubscript{2} is directly mixed in to cool the alcohol. Around $-$100$^{\circ}$C a very viscous slush is formed. As the LN\textsubscript{2} evaporates with time, the bath and the detector will warm up. Whenever necessary more LN\textsubscript{2} is added while stirring the mixture to keep it homogeneous. Still the temperature and pressure in the detector change continuously within a tight range. This range can be kept small, and the liquid in the detector is kept from freezing or boiling. The conditions in the detector are sufficiently stable to enable physics measurements, typically lasting a few hours.

This method is not practical since it requires a constant surveillance of the temperature and pressure in the detector, and a frequent intervention by stirring the alcohol slush. The method was never automated, and the monitoring proved rather distractive during experiments. There are also concerns regarding the safety, given the flammability of ethyl alcohol.

\subsection{LN\textsubscript{2} with Cooling Coil}
\label{sec: LN2 with Cooling Coil}

When the LXe is directly cooled as in the previous examples care must be taken that the cooling does not go below the freezing point. This is much easier if not the liquid is cooled, but the gas on top which will condense and accumulate in the vessel. One such indirect method involves a LN\textsubscript{2} cooling coil within the gas phase of the xenon. Of course at LN\textsubscript{2} temperature, a layer of frozen xenon will form around the windings of the coil. The xenon ice limits the heat flow since unlike the liquid the frozen xenon is stationary and there is no thermal transport by convection. Reducing the heat flow from the xenon to the coil naturally reduces the cooling action. The ice acts as an intrinsic self-regulating mechanism. Changing the LN\textsubscript{2} flow might cause too much or too little heat flow for the regulation to be effective.

A drawback of the cooling with a LN\textsubscript{2} coil is the varying xenon ice accumulation around the coil. The liquid level will not be constant and yet the liquid level stability is one of the requirements for the dual phase LXe detector, where electrons are extracted from the liquid to be observed in the gas on top.

\begin{figure}[htbp]
\centering % \begin{center}/\end{center} takes some additional vertical space
\includegraphics[width=.8\textwidth]{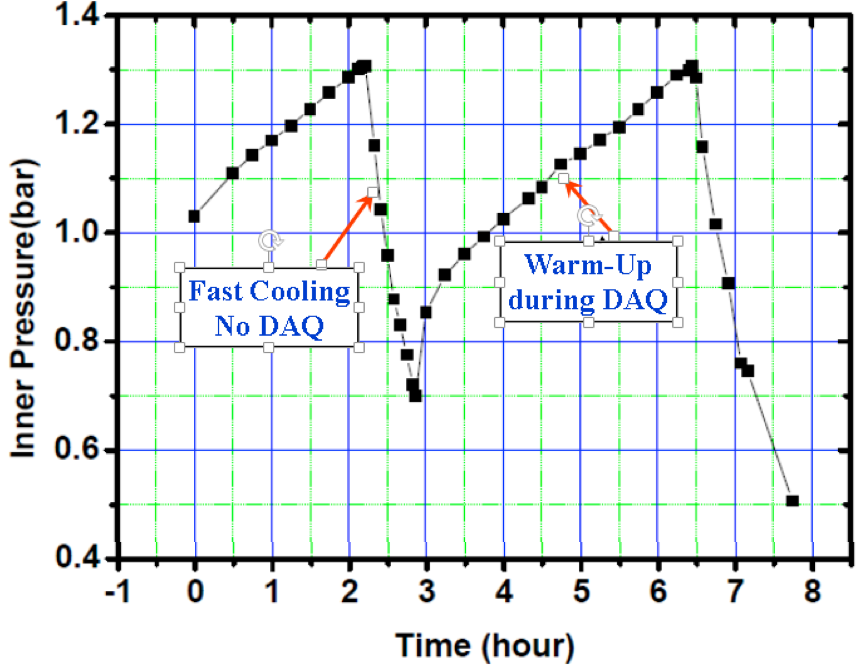}
\caption{\label{fig:1} Pressure variations in the PandaX detector~\cite{23} during cooling with a LN\textsubscript{2} cooling coil.}
\end{figure}

In principle the cooling power can be regulated by the flow of LN\textsubscript{2}. The xenon ice has a large heat capacity but the heat exchange with the gas is rather small. Therefore it takes long time until the frozen xenon liquefies again. It is difficult to find an equilibrium between the transported thermal energy and the required cooling power of the detector. Thus an appropriate operating point with a constant temperature is difficult to establish with a continuous LN\textsubscript{2} flow and would be time consuming in fine tuning of the system. Normally a simple regulation with two set points for the xenon gas pressure controls the LN\textsubscript{2} flow, e.g. 1.3 barG and 0.5 barG. The pressure in the detector thus constantly changes and follows a typical saw tooth structure shown in Figure~\ref{fig:1}. The method was successfully used on the high altitude balloon flights of the Liquid Xenon Gamma Ray Imaging Telescope (LXeGRIT)~\cite{18}, but data acquisition had to be stopped during the cooling cycles. The overall performance of the cooling system still was satisfactory and enabled measurements with this first LXe Time Projection Chamber (TPC) operated in near space environment. With a small 100 L LN\textsubscript{2} dewar carried on board, several flights in excess of 35 hours at 128,000 ft were achieved.

As mentioned above, the cooling coil is not very satisfactory to cool a DM detector during data taking. It is however a very simple and reliable way of refrigeration. The XENON and PandaX detectors use LN\textsubscript{2} coils for emergency cooling, e.g. in case of a power failure with no data taking.

\subsection{Pulse Tube Refrigerator}

\begin{figure}[htbp]
\centering % \begin{center}/\end{center} takes some additional vertical space
\includegraphics[width=.6\textwidth]{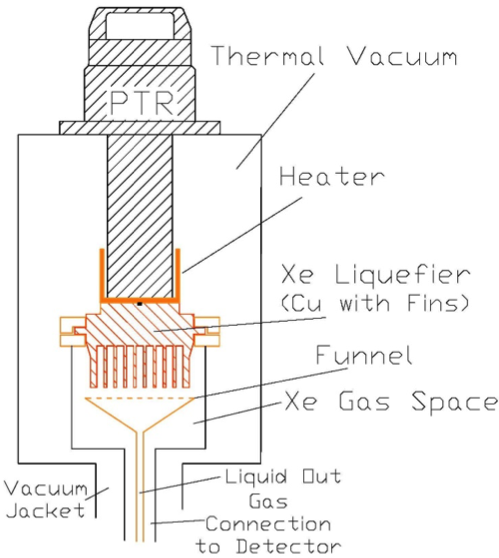}
\caption{\label{fig:2} Schematic drawing of the Cooling Module for a LXe set up. The unit is mounted some distance away from the detector (remote). The connection to the detector is via a Triple Line, liquid xenon in the innermost tube, surrounded by a gas xenon tube, surrounded by thermal vacuum.}
\end{figure}

All the XENON and PandaX DM experiments with dual phase xenon TPCs, as well as the single phase XMASS DM LXe detector~\cite{8}, have been cooled by a Pulse Tube Refrigerator (PTR)~\cite{19} specifically designed and optimized at KEK~\cite{20} for the use with LXe. A PTR is also employed for many LXe detectors built for laboratory setups by the above collaborations. From the first application on the XENON10 DM detector~\cite{9} developed at Columbia University, this cooling method has proven to be smooth, reliable, very stable, and easy to operate. The XENON10 detector used the first version of the PTR developed by the Iwatani Company, i.e. a P90 with 100 W cooling power. We concentrate our discussion on the subsequent and larger PTR, the Iwatani PC150, which was first used on the XENON100 DM detector~\cite{10}. Designed originally for 150 W cooling power, the PC150 can be boosted to 200 -- 250 W. For example, in the XENON1T experiment using a TPC filled with 3.2 t of LXe, the cooling power of the PC150 was measured as 250 W with a 7 kW compressor~\cite{21} at 50 Hz and 250 psi static helium pressure, once the PTR needle valve opening was optimized~\cite{1}. During operation the PTR always runs at its maximum strength. The cooling power is regulated by a heater unit on the cold head (see Figure~\ref{fig:2}). The heater is powered by a proportional-integral-derivative (PID) controller which delivers resistive heating based on the actual and set temperatures.

\begin{figure}[htbp]
\centering % \begin{center}/\end{center} takes some additional vertical space
\includegraphics[width=.8\textwidth]{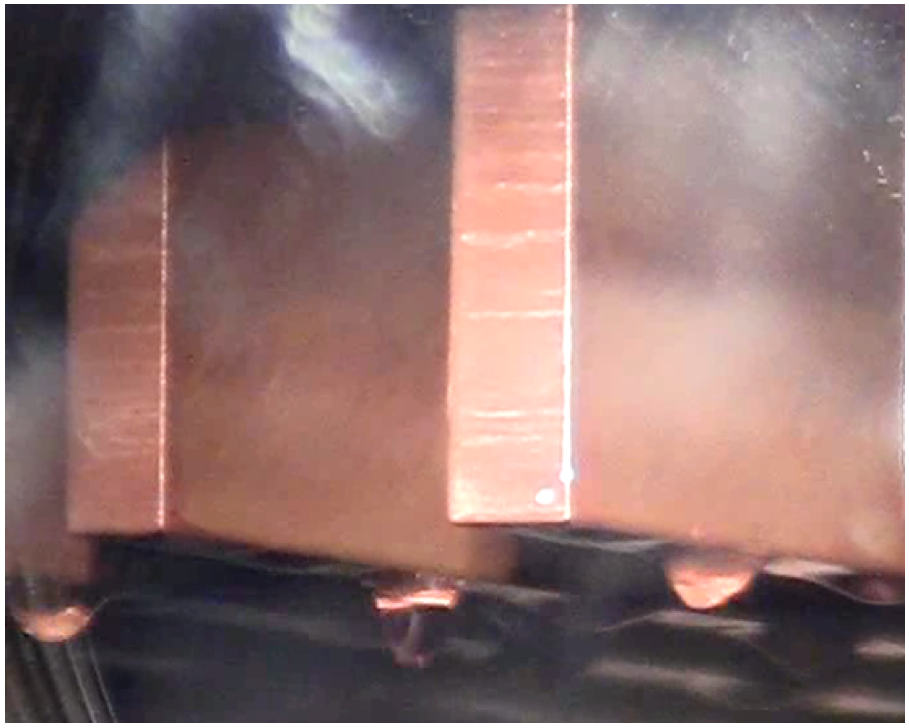}
\caption{\label{fig:3} LXe droplets forming on the copper fins cooled by a PTR. The fins are 1/4" wide. The diameter of the droplets is about 3/16".}
\end{figure}

The cold head of the PTR itself does not reach into the xenon filled space. Via a heater module, it connects to a copper cooling block penetrating the vessel wall. This deviates from the initial use of PTRs in the MEG experiment~\cite{22}. The indirect way of cooling was originally introduced in XENON10~\cite{9} to maintain the high purity requirement of the xenon target. The temperature sensors, the wires for the heater, and the feedthroughs are within the space of the thermal vacuum insulation, and thus do not contaminate the LXe. The cooling block is made of OFHC copper, and is hermetically sealed to the vessel walls. Only one end protrudes into the xenon, and here large fins result in better heat exchange with the gas. The seal is accomplished with an indium or aluminum wire. Xenon gas liquefies on the fins and droplets form (see Figure~\ref{fig:3}). They are collected, and the liquid is guided over a long distance to the detector, via an insulated line.

If more than 200 W cooling power is required, one can use two or more PTRs in parallel. This solution is implemented in the PandaX IV experiment. The increase of cooling power to 400 W comes with twice the power consumption for two helium compressors. Power consumption and costs do not favor designs with more than 2 PTRs in parallel.

For XENON1T~\cite{1}, a cooling system based on a single PTR was successfully implemented, with a second PTR serving as a backup unit to enable continuous operation in case of maintenance or failure of the main unit.

\begin{figure}[htbp]
\centering % \begin{center}/\end{center} takes some additional vertical space
\includegraphics[width=.8\textwidth]{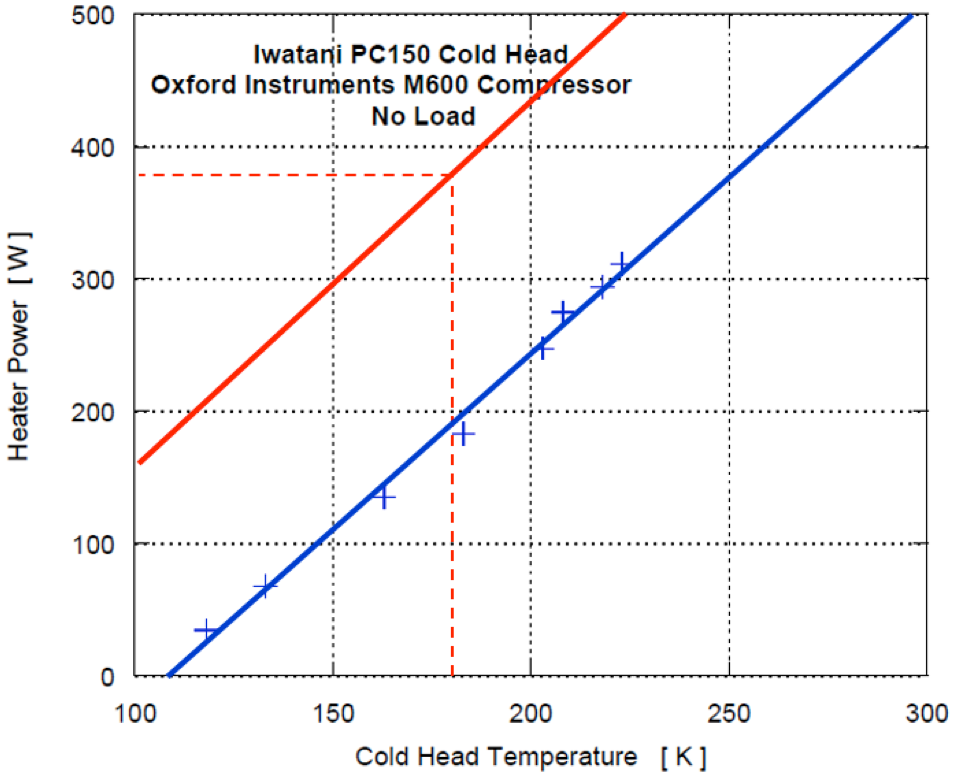}
\caption{\label{fig:4} Cooling power of the PTR versus the temperature of the cold head. The blue marks are measured data points with the PTR head at ambient temperature. The cooling power only depends on temperature differences. The red line is shifted by 70 degrees for a PTR with cooled head. From Ref.~\cite{23}.}
\end{figure}

A different approach to enhance the cooling power of a PTR proposes to cool the warm side of the PTR, the one exposed to the surrounding air. The cooling power of a given PTR is independent of the absolute temperature depending only on the temperature difference. Figure~\ref{fig:4} shows the measured cooling power vs. temperature~\cite{23}, with and without cooling of the PTR cold head top. The lowest temperature with no appreciable heat load is 110 K. At the LXe temperature of 173 K ($-$100$^{\circ}$C) the cooling power was measured as 180 W. With additional refrigeration of the cold head top to 223 K ($-$50$^{\circ}$C) the temperature difference the PTR had to provide was reduced from 120 degrees to 50 degrees. Under these conditions, the cooling power available was 380 W, twice the original value.

For temperatures around $-$50$^{\circ}$C one can use high power mechanical chillers at low cost and reasonable power consumption. However, we must note that one has to cool everything on the warm side of the PTR, and this includes the incoming helium stream on both sides, i.e. from the motor valve and from the buffer volume.

The described ways to boost the cooling power are economical and easy solutions in case only a modest increase is desired. They do not help, however, when a large factor is needed.

\subsection{LN\textsubscript{2} with Cold Finger}
\label{sec: LN2 with Cold Finger}

A LN\textsubscript{2} cooling with a cold finger is commonly used to improve the performance of a high resolution germanium (Ge) detector, by keeping the crystal at cryogenic temperature. The copper cold finger originates in a LN\textsubscript{2} reservoir with the other end directly connected to the Ge crystal. Since such detector type is a commercial product, the cooling design is highly optimized to keep the crystal at the best operating temperature, much warmer than LN\textsubscript{2}.

\begin{figure}[htbp]
\centering % \begin{center}/\end{center} takes some additional vertical space
\includegraphics[width=.6\textwidth]{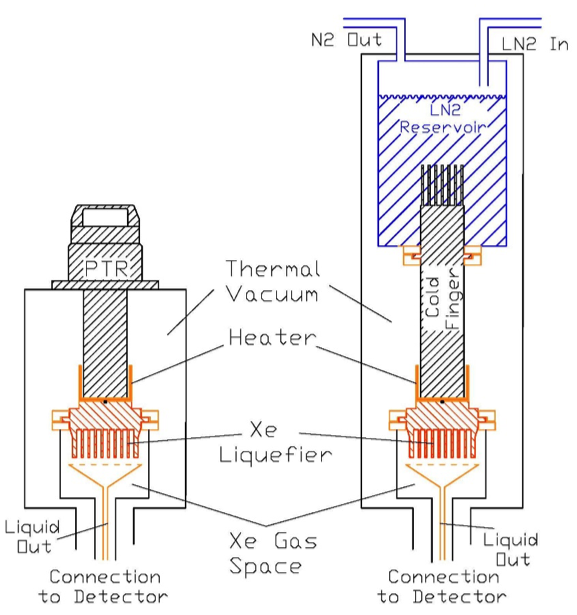}
\caption{\label{fig:5} Schematic comparison of a cooling module with PTR (left) and LN\textsubscript{2} cold finger (right). The similarities of the interface are obvious. The top of the cold finger also has fins for better heat transfer to the LN\textsubscript{2}.}
\end{figure}

In the case of a LXe detector however the heat load is not constant as it is affected by changes in gas recirculation flow and power dissipation of immersed light sensors, such as PMTs. The PTR can be replaced by a LN\textsubscript{2} reservoir with a cold finger (Figure~\ref{fig:5}), but one needs to control the supplied cooling power. In analogy to a PTR installation, we can introduce a heater plane at the end of the cold finger. The heater is a plane sheet of thick copper extending beyond the diameter of the cold finger. On the free surface high power resistors are mounted in good thermal contact with the copper plane. To dimension the cold finger we have to decide on the maximum cooling power, e.g. 2.5 kW, and we have to fix the lowest desired temperature, e.g. 163 K. The cold finger does have a thermal resistance, and the optimal cross section and length can be chosen such that at the maximum cooling load the lowest temperature can still be reached. Thus the cold finger shifts the temperature. If during operation less cooling power is required, or if the operating temperature is higher, the excess cooling power has to be dissipated by the heater module. Like the PTR, the LN\textsubscript{2} cold reservoir will always deliver the maximum cooling power. The heater neutralizes the superfluous cooling power. It can be controlled by a PID temperature controller, just like a PTR setup.

The reservoir should be a closed volume to avoid venting the boil-off nitrogen into the lab environment. The nitrogen gas is removed via an exhaust line. The reservoir can be automatically filled with LN\textsubscript{2} when the level falls below a set point. A cooling system with all the benefits of a PTR system but without the cooling power limitation is thus achieved. The system is also more economical.

\begin{figure}[htbp]
\centering % \begin{center}/\end{center} takes some additional vertical space
\includegraphics[width=.8\textwidth]{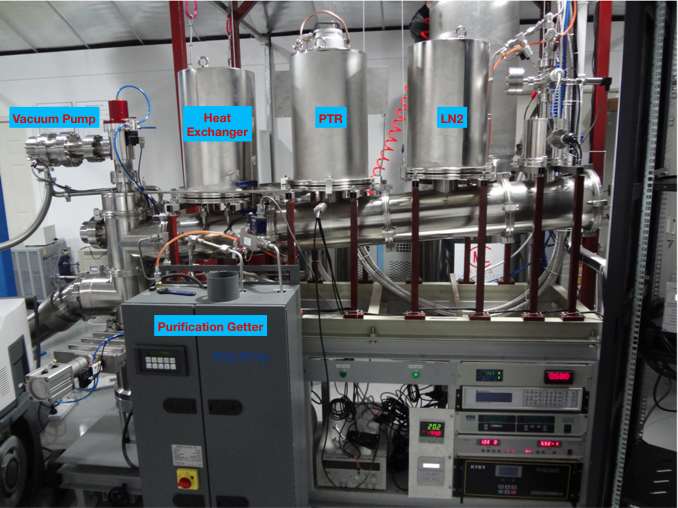}
\caption{\label{fig:6} The cooling system with a Cooling Bus structure. Each function is implemented in a separate module. The modules connect to the same tubes, the 'Bus'. The PTR module can be changed without affecting the other modules. The Bus is a Triple Line, liquid xenon in the innermost tube, surrounded by a gas xenon tube, surrounded by thermal vacuum.}
\end{figure}

The described LN\textsubscript{2} cooler can easily be integrated into an experiment if a modular architecture is used such as the 'Cooling Bus~\cite{23}' developed tor the PandaX experiments. A photo of the system used for PandaX I and II is shown in Figure~\ref{fig:6}. The present PTR module can simply be replaced by a LN\textsubscript{2} module. The Cooling Bus interfaces to the xenon gas space in the detector and is located outside the 5 m water shield above the detector. This principle of remote cooling was originally introduced for XENON100~\cite{10}.

The copper structure conducting the heat is obviously made of several pieces. They are bolted together for good thermal contact. The joining surfaces must be plane, better polished or lapped.

\section{Cooling a Very Large DM Experiment}
\label{sec: Cooling a Very Large DM Experiment}

\subsection{Cooling Power Requirement}

The required cooling power depends not only on the mass of the detector but also critically on other design parameters like the quality of the thermal insulation, the number of electrical feedthroughs, the thermal conduction in the mechanical support structure, etc. For any degree of accuracy a full thermal model of the detector is required. In lack of such a detailed study we can only estimate the cooling power, extrapolating from the experience with previous experiments, namely the XENON1T and PandaX II detectors. Detailed estimates for both experiments showed that a single PTR is sufficient during normal operation but this might limit high speed recirculation. We also have to remember that the heater control loop requires a certain margin for the regulation. Thus, the available cooling power should be at least 20\% above the peak consumption. During the filling of the detector the latent heat of the total xenon mass has to be provided additionally. The power requirement thus depends on the tolerable total filling time. There is a caveat, however. The cooling power during such peak times is normally very high, and the cooling module will provide this high power at all times. Practically this means that during normal operation with limited power requirements the difference has to be wasted with a heater. It might be better to design the system with two units, a low power unit for continuous cooling and a high power unit for peak times such as filling.

We discuss first the cooling requirements during normal operation. In the present experiments the cooling power provided by a PTR amounts to about 200 W, reduced by about 20\% as margin for the regulation. The load can be separated into cooling losses of the detector, losses of the structure itself, and the inefficiency of the heat exchanger used during recirculation.

Starting with the last term we have to fix the maximum recirculation speed. This is now limited to about 100 SLPM (Standard Liter per Minute) by the high temperature getter used in typical xenon gas purification systems. This speed is considered adequate during the initial phase of a data taking run, later to be reduced to about 30 SLPM when the liquid is sufficiently clean and only the residual outgassing has to be removed. Therefore we set the recirculation speed to 100 SLPM. The cooling and liquefaction of 1 SLPM of xenon from ambient temperature consumes 11 W. However an equal amount of heat has to be supplied to the xenon when boiling it off for purification in the gas phase. A heat exchanger~\cite{24} with >95\% efficiency can connect both processes. Thus the 1100 W cooling power for xenon liquefaction is reduced to 55 W, which is due to the heat exchanger inefficiency.

The second component is the heat loss in the cooling system itself including the long connecting tube from the cooling system to the detector. In terms of size and mechanical construction the cooling system and the connecting tube will not be very different from the present installations. The cooling power loss will be very similar. In total it is around 40 W. With better heat insulation one might even go below this value.

A last component is caused by the detector itself, and is governed by the thermal insulation of the vessel and its mechanical support. The thermal insulation can be improved significantly. Also the mechanical support and the connections to the inner vessel can be improved by either choosing materials with higher thermal resistivity, smaller cross sections, or longer effective thermal path lengths. The detector vessel is likely to be 16 times larger in volume than the present XENON1T detector. If we keep the same cylindrical shape of the LXe vessel, the heat transport does not scale with the volume or mass of the detector, but with the surface. This means for the 16 times larger volume we need about 6.3 times the cooling power. We not only have the conduction via the vessel wall but also the conductive heat transport via the vessel support. It would scale with the mass, but there are several ways to reduce the thermal conductance of the supporting elements.

This component also includes heat entering via electrical connections, mainly the high voltage (HV) connection for the cathode and the cables for the PMTs, in the case of typical LXe detectors used for DM search. With the increased number of PMTs this heat transfer becomes significant. One also has to add the electrical power dissipated in the PMT base circuits which is of order 22 mW per PMT. Considering a total of 3000 PMTs, the resulting 70 W of electrical power is not negligible. This heat, however, is localized on the resistors of the PMT bases. Thus it might lead to the formation of bubbles, which must be avoided. Eventually, the base circuits have to be modified to significantly reduce this resistive heating. This still leaves the heat conduction through HV and signal cables of the PMTs. The easiest way to reduce this term is to convert from coaxial cables to balanced Kapton strip lines~\cite{25}.

The total heat load to the system can thus be written as:
\begin{equation*}
\label{eq:1}
W_{tot} = W_{CB} + W_{HE} + W_{PMT} + W_{CAB} + W_{V}
\end{equation*}
where \(W_{tot}\) stands for the total heat load, \(W_{CB}\), \(W_{HE}\), \(W_{PMT}\), \(W_{CAB}\), \(W_{V}\) denote the heat load from the cooling system, recirculation, PMT bases, cabling, and vessel walls respectively.

As discussed before the \(W_{CB}\), and the \(W_{HE}\) terms do not depend on the detector mass but only on the cooling and the recirculation system. There are even simple design changes to improve the thermal insulation. \(W_{PMT}\) would become significant because of the much larger number of PMTs. But we also run the risk of local heating and bubbling. A change of the PMT bases will be required to reduce the dissipated power. As previously said, switching from coaxial cables to flat cables will help to reduce this heat load. \(W_{CAB}\) will be reduced at the same time, since flat cables have a much lower heat transport by eliminating the shield of coaxial cables. A design goal should be an overall PMT power less than 10 W. The only term which changes our balance of cooling power is the \(W_{V}\) term from the vessel. It depends critically on the actual design, i.e. the thermal insulation and the materials used. Assuming a similar design as present DM detectors, the value would not change with the volume or the xenon mass, but again with the surface area, i.e. a factor of 6.3 instead of 16.

Let's use some numbers from the experience with current detectors, such as PandaX. The cooling system including the connecting tube should be around 40 W. The recirculation with 100 SLPM stays the same with 60 W. The PMT bases when modified remain below 10 W. The heat transfer through the vessel walls and the conduction through the support is now about 50 W. For a current system, we reach 160 W plus 20\% for the regulation, still in the range of a single PTR. A 50 t detector, about 16 times larger than the present XENON1T, will increase \(W_{V}\), but with a careful design we should stay below a factor of 6.3, adding 300 W. If we include a 20\% margin for regulation, we end up with a total just short of 1 kW.

The above estimate is only valid during normal operation, i.e. data taking. The exact value can only be derived once the actual size of the detector is fixed. Also the design of the thermal shielding can change the estimate in both directions. Once all this information is available a detailed thermal analysis can predict a more accurate value of the required cooling power.

\subsection{Filling Procedure}

During filling from the gas phase we cannot engage the heat exchanger. If we assume a total mass of 50 t we have to process $9.1 \times 10^{6}$ standard liter equivalent xenon (5.5 g/standard liter). We know from the recirculation design that we need 11 W to cool and liquefy at a rate of 1 SLPM. Now we have to decide on a reasonable filling time. We assume the xenon is stored as gas and we want to pass it through the getter before entering the detector, i.e. the flow rate is limited to 100 SLPM. We need 64 days for this process with a continuous cooling power of 1100 W, or 1690 kWh.

To reduce this long filling time we can convert to filling in the liquid phase. In this case we need a liquid xenon storage vessel above the water shield. The storage should be a double walled steel vessel with a very good thermal vacuum insulation. Initially the xenon is liquefied into this vessel. But the filling still needs 1690 kWh. If there is not enough time we shall use a LN\textsubscript{2} coil in good thermal contact with the vessel wall to freeze the xenon. Once liquefied it will be stored while passing through the recirculation-purification system. During the filling the xenon flows in liquid phase propelled by gravity into the pre-cooled detector. Now we can fill at any rate just offsetting the thermal losses in the transfer tube.

The advantage of liquid filling stems from the idea that the detector is only emptied to service the inner part, namely the TPC structure. But this too requires a lot of time. The xenon has to be stored during this period in a storage tank. The recirculation system can keep the liquid clean. Once the service to the TPC is finished, and the vessel is closed again, we can fill the detector within very short times, maybe on the order of a few days. A similar system was used for the MEG experiment~\cite{22}.

\subsection{Cold Finger Design}

The cold finger connecting the LN\textsubscript{2} reservoir to the detector is simply a thick copper rod of 1" to 2" diameter. It is dimensioned such that the lowest desired temperature can still be reached at the maximum desired cooling power. This temperature must be above the freezing point of xenon so that at no time can any xenon solidify. The copper rod just shifts the temperature from the LN\textsubscript{2} value to the LXe range. We can easily derive the appropriate dimensions of the rod with the following calculation:
\begin{equation*}
\label{eq:2}
Q = k / l \times A \times \Delta T
\end{equation*}
where \(Q\), \(k\), \(l\), \(A\), \(\Delta T\) denote the heat load, thermal conductivity of copper (400 W/m\(\cdot\)K), length of copper rod, cross section of rod and temperature difference.

\(\Delta T\) is the difference between the LN\textsubscript{2} temperature ($-$196$^{\circ}$C) and operating temperature of the LXe detector ($-$100$^{\circ}$C). \(Q\) is then the design value of the cooling power for this module. The cold finger acts like a resistor in an electrical circuit reducing the potential. This can be split into the thermal resistance of the cold finger material itself and a term describing the condensation heat transfer which depends on the area, but also on the convection of the gas.

\subsection{LN\textsubscript{2} Consumption}

Almost all major labs have an economic supply of LN\textsubscript{2} in large quantities for cooling purposes. From there it can be delivered to the experiment either in movable dewars with 200 -- 250 L content, or via a fixed installation with a double walled insulated pipe. LN\textsubscript{2} for cooling is thus very convenient, and it is economically viable despite the elevated consumption for a massive LXe experiment.

The LN\textsubscript{2} consumption can be calculated from the latent heat of LN\textsubscript{2} evaporation. The value is 199 kJ/kg and the density is 0.8 kg/L. For normal operation with a 1 kW cooler we will use about 500 -- 550 L/day. Filling the storage vessel with a power of 2.5 kW requires 1250 L/day. The total time of liquefaction would be 28 days. The calculations do not include losses due to imperfect insulation of the connecting lines, etc.

The cost of LN\textsubscript{2} varies not only with consumption, but also with location. It is common to assume an average of \$0.2 /L. The 1 kW cooler would run for \$50 /day. As comparison, electrical energy has an average price of \$0.1 /kWh. A single 10 kW helium compressor for a PTR consumes the same amount if we include the costs for water chillers and additional air conditioning, etc. As for reliability, the LN\textsubscript{2} system does not have any moving parts. The PTR on the other hand needs the motor valve and the compressor, both requiring periodical servicing.

\subsection{Emergency cooling}

Now that the cooling is provided by LN\textsubscript{2}, do we still need an emergency cooling module? Without the helium compressor for the PTR the electrical power consumption is dramatically reduced. Since also the recirculation uses considerable power, it might be necessary to stop this, too. Of course, data taking also discontinues. With no need to keep the system in a tightly controlled equilibrium it is much simpler to control the emergency cooling by the pressure instead of the temperature. The essential parts for the cooling are reduced to the solenoid valves and the pressure controller. This amount of power can easily be supplied by a small Uninterruptible Power Supply (UPS) for a very long time. We thus could give up the emergency cooling module, but, there is no real benefit in doing so. The emergency module with a LN\textsubscript{2} coil is very easy to design, and the operation with a two set point controller is very simple and highly reliable. It adds some redundancy to the system and can keep the detector ready in case the main cooling system needs servicing. As an example, the XENON100~\cite{10} cooling system included such emergency cooling which turned out to be very useful throughout the multi-year operation of that experiment.

The backup LN\textsubscript{2} cooling of the XENON1T experiment was designed to offer additional features and thus is more complex. It uses a cold finger coupled to a LN\textsubscript{2} evaporator instead of the LN\textsubscript{2} cooling coil method described in Sec.~\ref{sec: LN2 with Cooling Coil}. The evaporator of the LN\textsubscript{2} cooling system is a pressure vessel whose bottom surface is thermally coupled to the cold finger. It is connected to a source of LN\textsubscript{2} at its inlet, and has a GN\textsubscript{2} outlet. The temperature of the cold finger is kept at the desired value by adjusting the rate at which LN\textsubscript{2} evaporates. This is accomplished through the use of a proportional control valve that limits the flow of GN\textsubscript{2} out of the evaporator. The control valve's opening is set by a PID controller that takes the cold finger temperature as input. The operation of the system requires less than 30 W of electrical power and achieves a temperature stability of about $\pm$0.5 K. In the case of a complete loss of electrical power, a pre-adjusted needle valve sets the GN\textsubscript{2} flow, and thus the cooling power, to a safe level.

\subsection{R\&D and Tests with a LN\textsubscript{2} Cooling System}
\label{sec: R&D and Tests with a LN2 Cooling System}

The proposed cooling method was previously used for tests with LXe detectors at Columbia University, but unfortunately it was never described in a publication. One such unit was designed in 2004 for a LXe chamber of about 3 liters volume. The LN\textsubscript{2} - alcohol cooling option was not well suited since the experiment~\cite{26} was located within the limited access area of a neutron test beam. The necessary frequent interventions to monitor the cooling mixture and to add LN\textsubscript{2} would have been intolerable. The LN\textsubscript{2} was provided by a 100 L reservoir, and a simple two set point control circuit kept it filled for many hours of uninterrupted data taking. The cooling proved to be very convenient and reliable, approaching the convenience of a PTR system. Later on the detector was modified and used for diverse studies~\cite{27} in the lab.

\begin{figure}[htbp]
\centering % \begin{center}/\end{center} takes some additional vertical space
\includegraphics[width=.8\textwidth]{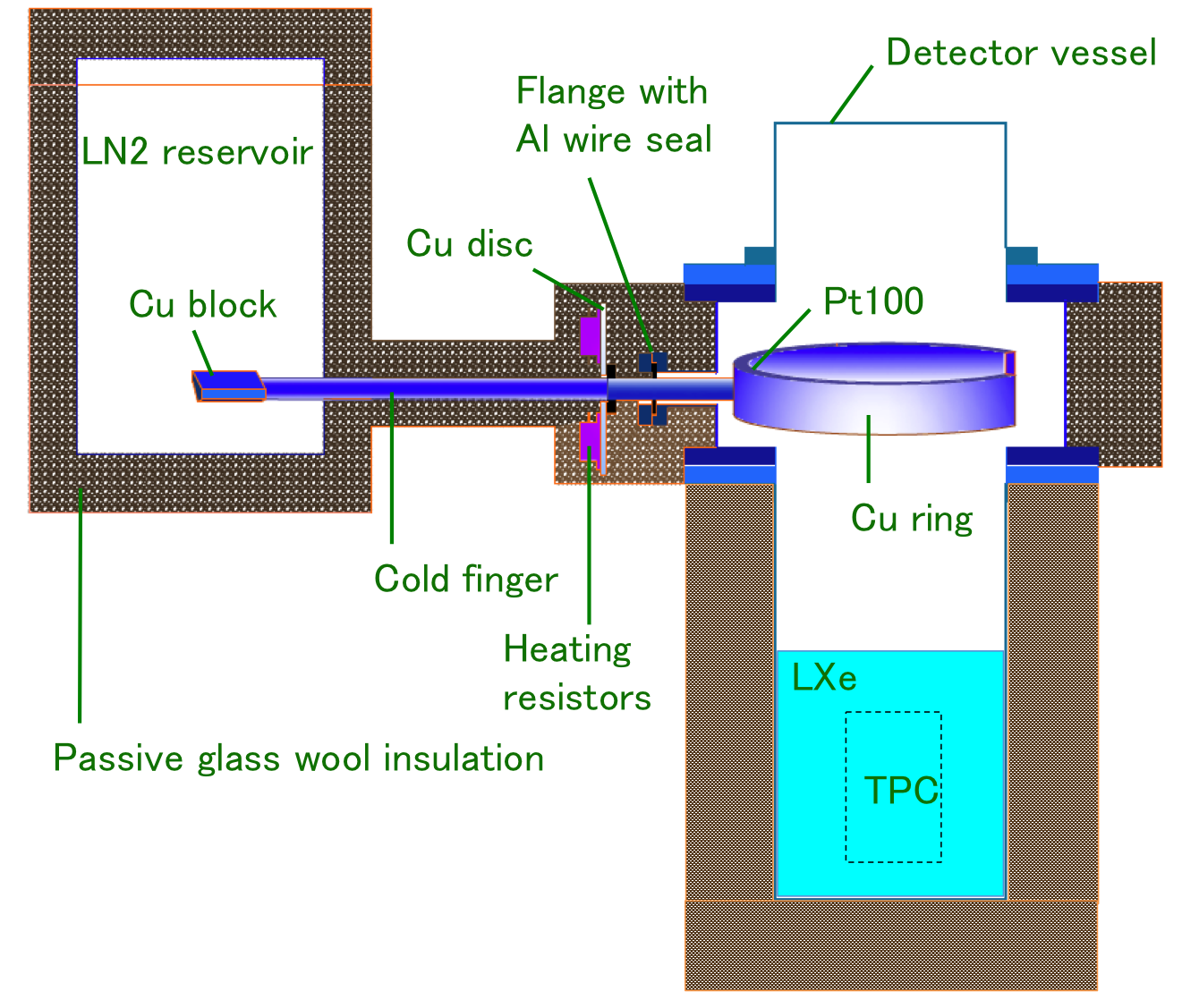}
\caption{\label{fig:7} Schematic view of the cooling system used in Ref.~\cite{27}. The end of the cold finger is a 1/4" thick copper ring. Thus the center is free for electrical connection from the top flange.}
\end{figure}

The design aimed at a very economical, yet reliable and efficient cooling. Instead of the usual vacuum insulation, passive insulation was used both for the LN\textsubscript{2} reservoir and the detector itself. Although functional, the insulation proved less than optimal and unnecessarily enhanced the cooling power requirements. Figure~\ref{fig:7} shows a schematic view of the system.

The cold finger penetrates the reservoir wall on one end and the chamber vessel at the other. After entering the detector it ends in a large 1/4" thick and 1" wide copper ring for good thermal exchange with the xenon gas. A copper disk 1/4" thick and 5" diameter is mounted in the path of the cold finger as heater unit. A series of high power resistors are bolted to the disk. On the other side the 1.25" diameter cold finger rod reaches far into the reservoir. The steel to copper interfaces on the walls are sealed with indium wire in a V-groove. A Pt-100 temperature sensor after the heating element completes the thermal system. Despite the poor performance of the passive insulation, the cooling power of the cold finger was still excessive for the output of our PID controller of 100 W. Half of the resistors were therefore powered with a constant DC power supply.

The described R\&D system developed at Columbia University contains all the ingredients for a future DM detector cooling system based on LN\textsubscript{2} as proposed here. Of course it is for xenon, i.e. the temperature range is the same, it is PID-controlled, it can have a very high cooling power output depending on the cold finger dimensions, it cools indirectly, i.e. only the end of the cold finger comes in contact with xenon, and it can cool remotely, although this function was not implemented in the R\&D setup. Thus, the 'remote cooling' was reduced to zero-length since an additional vessel at some distance would have unnecessarily complicated the system.

\section{Conclusion}
\label{sec: Conclusion}

In astroparticle physics research massive LXe detectors, well above the 10 t range of those currently being developed, are surely on the horizon. They require much more cooling power, although better thermal insulation and a better thermal design can alleviate the problem. A simple projection puts the cooling power demand at 1 -- 1.5 kW for a 50 t detector. This is for normal operation only. The xenon filling phase would require 2.5 kW or more for an acceptable filling period.

The presently popular PTR based xenon cooling system is very convenient for operation. It has proven to be very reliable with good short and long term stability. However, it will not be able to provide the large amount of cooling power required in the future. Further drawbacks include the high costs of these units and the high electrical power consumption for the helium compressor, which of course generates a lot of heat and mechanical noise. In an attempt to maintain the attractive benefits and simple operation of a PTR system as much as possible we propose to replace the cooling unit by a LN\textsubscript{2} reservoir with a cold finger.

The change of cooling method is not very complex. In a modular architecture like the previously proposed Cooling Bus of the PandaX experiment the design of the new cryogenic system is reduced to a direct replacement of the cooling unit. The other functions and the remaining modules are not affected. A small system employing this cooling method was designed and tested at Columbia with excellent results.

\acknowledgments

We would like to acknowledge the ideas and the help from many members of the XENON and PandaX teams. Their assistance and contributions were very important for the successful design, operation, and the tests with the small detector system. Especially many discussions with G. Plante were instrumental for this publication.

This project has been supported by a grant from the Ministry of Science and Technology of China (Grant NO.2016YFA0400301). At Columbia and Rice University, the work was supported by grants for the XENON Dark Matter Project.

\end{document}